\documentclass[twocolumn,prl,aps,showpacs]{revtex4}
\usepackage{epsfig,psfrag}
\usepackage{latexsym}

\newcommand{\be}{\begin{equation}}
\newcommand{\ee}{\end{equation}}

\begin{document}
\title{Heat fluctuations in Ising models coupled with   two different heat baths}

\author{A. Piscitelli}
\email{antonio.piscitelli@ba.infn.it} \affiliation{Dipartimento di Fisica, Universit\`a di
Bari {\rm and} Istituto Nazionale di Fisica Nucleare, Sezione di  Bari, via Amendola 173,
70126 Bari, Italy}
 \author{F. Corberi} \email{federico.corberi@sa.infn.it}
 \affiliation{Dipartimento di Matematica ed Informatica,
via Ponte don Melillo, Universit\`a di Salerno, 84084 Fisciano (SA), Italy}
  \author{G. Gonnella}
\email{giuseppe.gonnella@ba.infn.it} \affiliation{Dipartimento di Fisica, Universit\`a di Bari
{\rm and} Istituto Nazionale di Fisica Nucleare, Sezione di  Bari, via Amendola 173, 70126
Bari, Italy}

\date{\today}
\begin{abstract}
Monte Carlo simulations of Ising models coupled to  heat baths at two different temperatures
are used  to study a fluctuation relation   for the heat exchanged between the two thermostats
in a time $\tau$. Different kinetics (single--spin--flip or spin--exchange Kawasaki dynamics),
transition rates (Glauber or Metropolis), and couplings  between the system and the
thermostats have been considered. In every case the fluctuation relation  is verified in the
large $\tau $ limit, both in the disordered and in the low temperature phase. Finite-$\tau$
corrections are shown to obey a scaling behavior.

\end{abstract}
 \vskip -0.2cm
\pacs{05.70.Ln; 05.40.-a; 75.40.Gb} \maketitle

In equilibrium statistical mechanics expressions for the probability of different microstates,
 such as the Gibbs weight in the canonical ensemble, are the starting point of a successful
theory which allows the description of a broad class of systems. A key-point of this approach
is its generality. Specific aspects, such as, for instance, the kinetic rules or the details
of the interactions with the external reservoirs are irrelevant for the properties of the
equilibrium state.

In non-equilibrium systems general expressions for  probability distributions are not
available; however,
the recent proposal~\cite{ECM93,ES94,GC95} of relations  governing the fluctuations is of
great interest. They were formalized, for a class of dynamical systems, as a theorem for the
entropy production in stationary states \cite{GC95}.
 Fluctuation Relations (FRs) have been established  successively for a broad class of
stochastic and deterministic systems
\cite{Jarzynski,Kurchan,LS,maes,Crooks,HatanoSasa,VZC,seifert}
 (see
\cite{rev} for recent reviews).
 They
  are expected
to be relevant
 in nano-- and biological sciences \cite{Bustamante} at scales where
typical thermal fluctuations are of the same magnitude of the external drivings. Testing the
generality of FRs and the mechanisms of their occurrence in experiments \cite{wang,cili} or
numerical simulations, particularly for interacting systems, is therefore an important issue
in  basic  statistical mechanics and  applications.

In this work we consider the case of non equilibrium steady states of systems
 in contact with two different heat baths at temperatures $T_n$ $(n=1,2) $. In this case the
FR, also known as
 Gallavotti-Cohen relation \cite{GC95},  connects the probability
$\mathcal{P}(\mathcal{Q}^{(n)}(\tau))$ to exchange the heat $Q^{(n)}(\tau)$ with the $n$-th
reservoir in a time interval $\tau$, to that of exchanging the opposite quantity
$-Q^{(n)}({\tau})$, according to
\begin{equation}
 \ln{\frac{\mathcal{P}(\mathcal{Q}^{(n)}(\tau))} {\mathcal{P}(-\mathcal{Q}^{(n)}(\tau))}} =
\mathcal{Q}^{(n)}(\tau) \Delta \beta^{(n)},
\label{FR}
\end{equation}
where $\Delta \beta ^{(1)}=
\left(\frac{1}{T_{2}} - \frac {1}{T_1} \right)$ and $\Delta \beta ^{(2)}=-\Delta \beta ^{(1)}$.
This relation is expected to hold in the large $\tau $ limit; in particular
$\tau $ must be much larger than the relaxation times in the system. An explicit derivation of
(\ref{FR}) for stochastic systems can be found in \cite{BD}. In specific systems, the
validity of (\ref{FR}) was  shown   for a chain of oscillators~\cite{lepri} coupled at the
extremities  to two thermostats while the case of a brownian
particle in contact with two reservoirs has been studied in ~\cite{visco}. A relation similar
to (\ref{FR}) has  been also proved for the heat exchanged between two systems initially
prepared in equilibrium at different temperatures and later put in contact~\cite{JW07}.

The purpose of this Letter is to study the relation~(\ref{FR}), and the pre-asymptotic
corrections at finite $\tau$,  in Ising models in contact with two reservoirs, as a
paradigmatic example of statistical systems with
 phase transitions. This issue was considered analytically in a mean
field approximation in~\cite{LRW} where the distributions
$\mathcal{P}(\mathcal{Q}^{(n)}({\tau}))$ have been  explicitly computed in the large $\tau $
limit. Here we study numerically the model with short range
interactions. This allows us to analyze the generality of the FR~(\ref{FR})
with respect to details of the kinetics and of the interactions with the
reservoirs,  and to study the effects of
finite $\tau $. We also investigate the interplay between ergodicity breaking and the FRs.

We consider a two--dimensional Ising model defined by the Hamiltonian $ H = - J \sum_{\langle
ij \rangle} \sigma_i \sigma_j$, where $\sigma_i = \pm 1$ is a spin variable on a site $i$ of a
rectangular lattice  with $N=M \times L$ sites, and the sum is over all pairs $\langle ij
\rangle $ of nearest neighbors. A generic evolution of the system is given by the sequence of
configurations $\{\sigma _1(t), \dots , \sigma _N (t)\}$ where $\sigma _i(t)$ is the value of
the spin variable at time $t$. We will study the behavior of the system in the stationary
state.  In order to see the effects  of different kinetics, Montecarlo single--spin--flip and
spin-exchange (Kawasaki) dynamics have been considered, corresponding to systems with
non-conserved or conserved magnetization, respectively. Metropolis transition rates have been
used; for single--spin--flip dynamics we also used Glauber transition rates.

Regarding the interactions with the reservoirs we have considered two different
implementations. In the first, the system is {\it statically } divided into two halves. The
left part (the first $M/2$ vertical lines) of the system interacts with the heat bath at
temperature $T_1$ while the right part is in contact  with the reservoir at $T_2>T_1$. We have
used both open or periodic boundary conditions. In the second implementation (for
single--spin--flip only) each spin $\sigma _i$, at a given time $t$, is put {\it dynamically }
in contact with one or the other reservoir depending on the (time dependent) value of
$h_i=(1/2)\vert \sum _{\langle j\rangle _i}\sigma _j \vert$, where the sum runs over the nearest
neighbor spins $\sigma _j$ of $\sigma _i$. Notice that $h_i$ is one half of the (absolute
value) of the local field. In two dimensions, with periodic boundary conditions, the possible
values of $h_i$ are $h_i=0,1,2$. At each time, spins with $h_i=1$ are connected to the bath at
$T=T_1$ and those with $h_i=2$ with the reservoir at $T=T_2$. Namely, when a particular spin
$\sigma _i$ is updated, the temperature $T_1$ or $T_2$ is entered into the transition rate
according to the value of $h_i$. Loose spins with $h_i=0$ can flip back and forth regardless
of temperature because these moves do not change the energy of the system. Then, as in the
usual Ising model, they are associated to a temperature independent transition rate (equal to
$1/2$ or $1$ for Glauber or Metropolis transition rates). This model was introduced
in~\cite{olivera} and further studied in~\cite{godreche}. It is characterized by a line of critical
points in the plane $T_1,T_2$, separating a ferromagnetic from a paramagnetic phase
analogously to the equilibrium Ising model.

Denoting with $t^{(n)}_k $
the times at which an elementary move is attempted by coupling
the system to the $n$-th reservoir,
the heat released by the bath
in a time window $t\subset [s,s+\tau]$ is defined as
\begin{equation}
\mathcal{Q}^{(n)}(\tau) = \sum_{\{t^{(n)}_k \} \subset [s, s+\tau]}
[H(\sigma(t^{(n)}_k)) - H(\sigma(t^{(n)}_k-1))].
\label{calore}
\end{equation}
In the stationary state, the properties of
$\mathcal{Q}^{(n)}(\tau)$
will be computed by collecting the statistics
over different sub-trajectories
obtained by dividing a long history of length $t_F$ into many ($t_F/\tau$)
time-windows of length $\tau $, starting from
different $s$.

\vspace{0.5cm}

$\mathbf {T_1,T_2>T_c}$. We begin our analysis with the study  of the relation~(\ref{FR})
in the case with  both temperatures well above the
critical value $T_c\simeq 2.269$ of the equilibrium Ising model.
In the following we will measure times in montecarlo steps (MCS)
(1 MCS=N elementary moves).

\begin{figure}
\includegraphics*[height=6.5cm,width=9.cm]{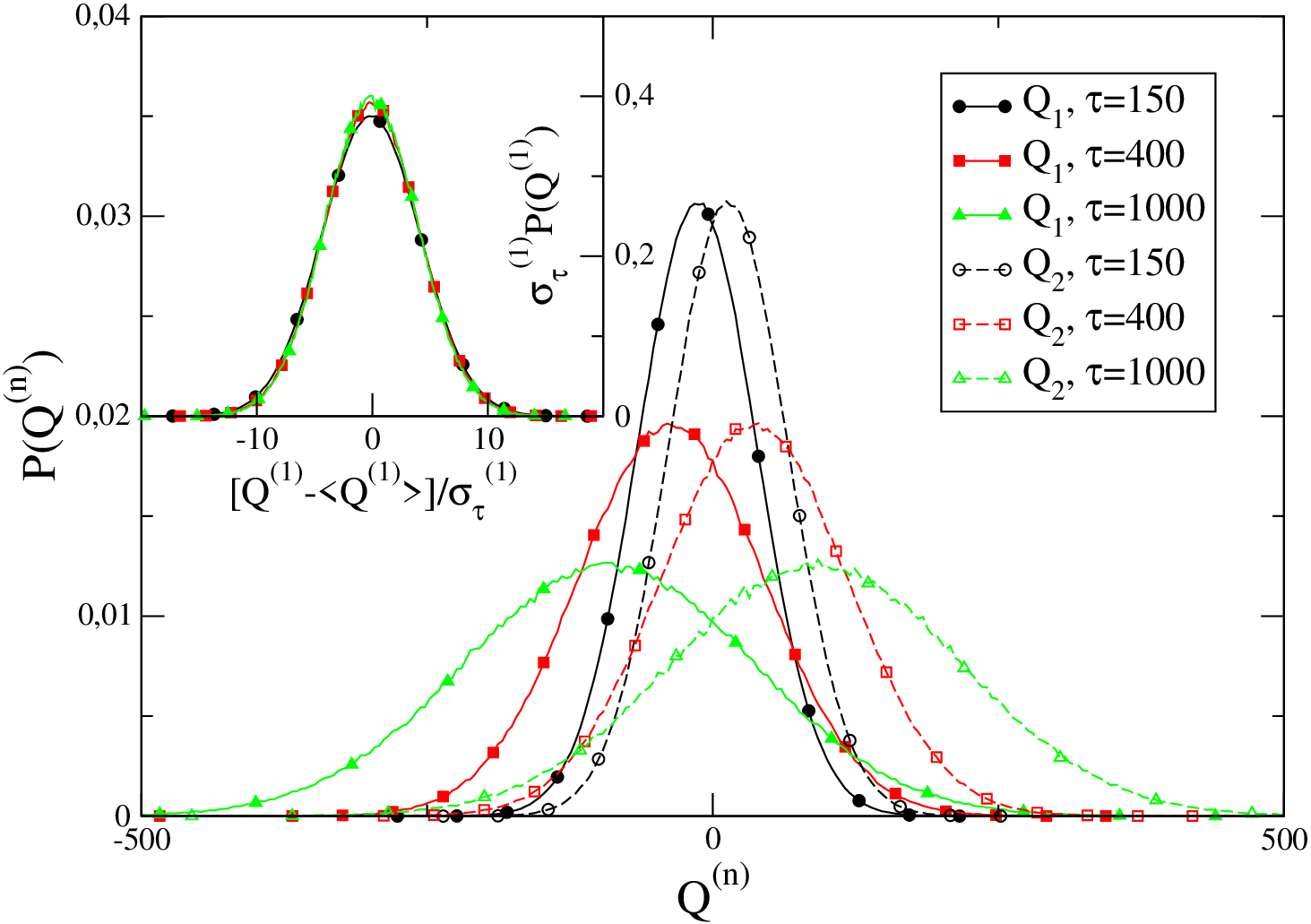}
\vskip 0.4cm
\includegraphics*[height=6.5cm,width=9.cm]{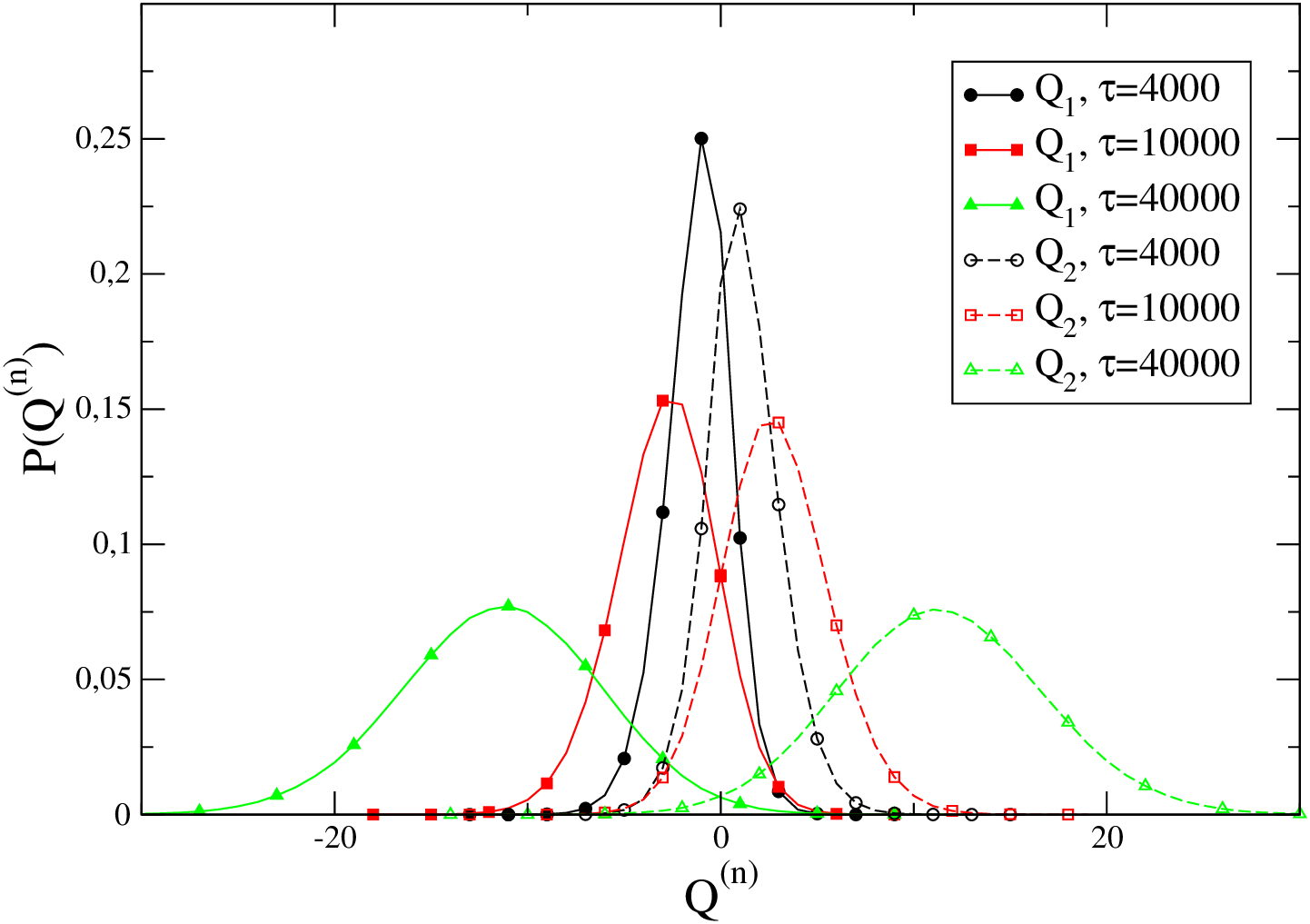}
\vskip 1cm
 \caption{Upper panel ($T_1,T_2>T_c$): Heat PDs for $\mathcal{Q}^{(1)}$ (on the left) and
 $\mathcal{Q}^{(2)}$ (on the right) for a system with $T_1 =2.9, T_2=3$, size 10x10 and
$t_F=6\cdot 10^8$ MCS. In the inset
$\sigma^{(1)}_{\tau}\mathcal{P}(\mathcal{Q}^{(1)}({\tau}))$ is plotted against
$(\mathcal{Q}^{(1)}(\tau) - \langle \mathcal{Q}^{(1)}(\tau)\rangle)/\sigma^{(1)}_{\tau}$.
Curves for different $\tau $ collapse on a Gaussian mastercurve. Lower panel ($T_1,T_2<T_c$):
Same kind of plot for $T_1 =1, T_2=1.3$, size 10 x 10 and $t_F= 10^9 $ MCS. The skewness of
the distributions (see text) is equal to 2.35, 0.589, 0.11 for the cases $\tau= 4000, 10000,
40000$ respectively.}

\label{pdfsaboveTc}
\end{figure}

\begin{figure}
\includegraphics*[height=6cm,width=8.cm]{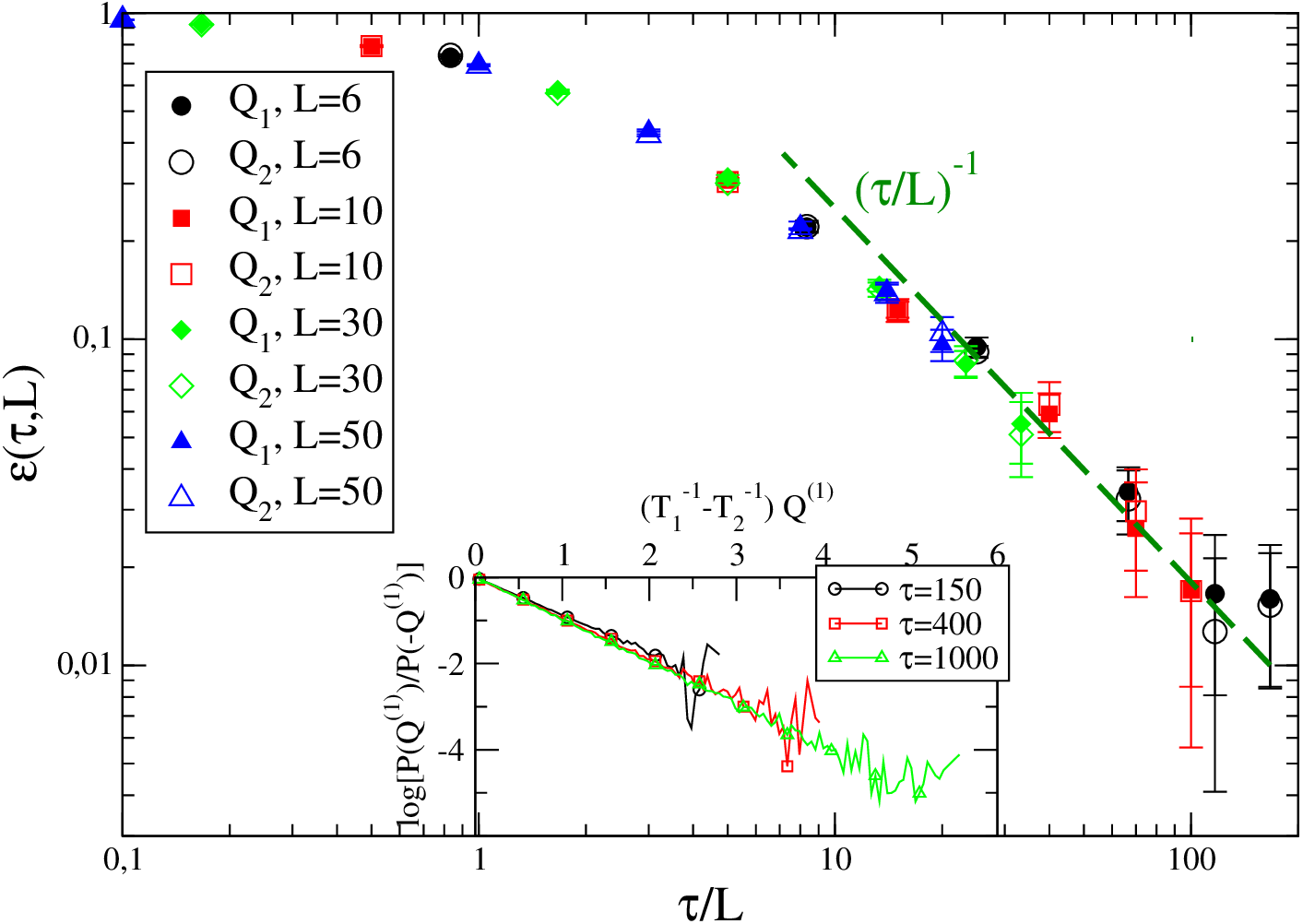}
\includegraphics*[height=6cm,width=8.5cm]{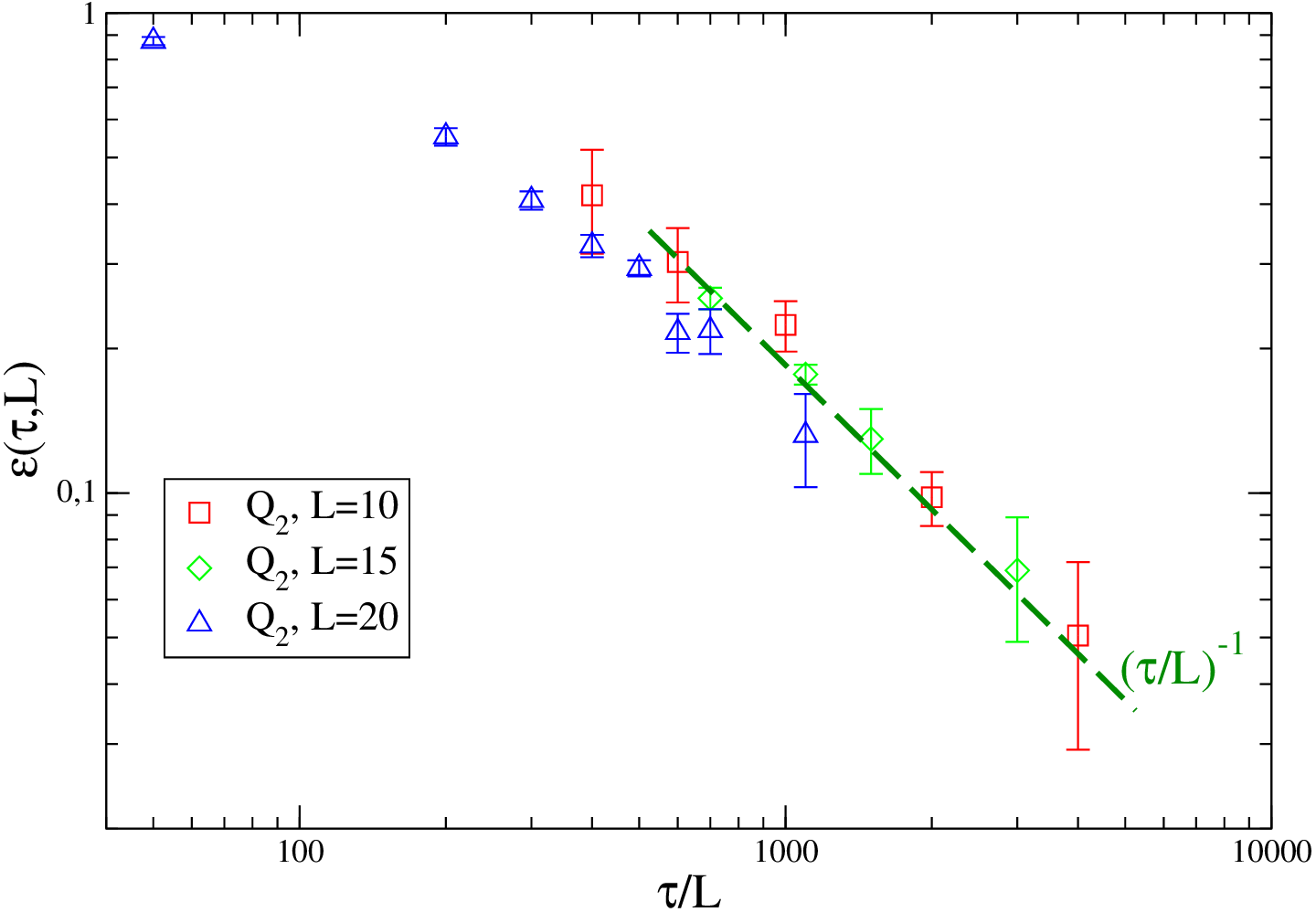}
\caption{Same parameters of Fig.~\ref{pdfsaboveTc} (different sizes).
Upper Panel ($T_1,T_2>T_c$): $\epsilon (\tau,L)$
is plotted against $\tau /L$ for different $L$. Inset: $\log \left
[\mathcal{P}(\mathcal{Q}^{(1)}(\tau))/ \mathcal{P}(-\mathcal{Q}^{(1)}(\tau))\right ]$ is
plotted against $(1/T_1-1/T_2)\mathcal{Q}^{(1)}(\tau)$.  Lower panel ($T_1,T_2<T_c$):
$\epsilon^{(2)}$ is plotted against $\tau /L$ for different $L$.} \label{rette}
\end{figure}

The typical behavior of the heat probability distribution (PD) is reported in the upper panel
of Fig.\ref{pdfsaboveTc} for the case with static coupling to the baths, single--spin--flip
with Metropolis transition rates, $T_1=2.9$, $T_2=3$, and a square geometry with $L=M=10$
(Much larger sizes are not suitable because trajectories with a heat of opposite sign with
respect to the average value would be too rare). As expected, $\mathcal{Q}^{(1)}(\tau)$
($\mathcal{Q}^{(2)}(\tau)$) is on average negative (positive) and the relation
$<\mathcal{Q}^{(1)}(\tau)> + <\mathcal{Q}^{(2)}(\tau)> = 0 $ is verified. Regarding the shape
of the PD, due to the central limit theorem, one expects a gaussian behavior for $\tau $
greater than the (microscopic) relaxation time (in  this case it is of few MCS ($\sim 5$)),
namely $\mathcal{P}(\mathcal{Q}^{(n)}({\tau}))= (2\pi )^{-1/2}(\sigma^{(n)}_{\tau})^{-1} \exp [-
\frac {(\mathcal{Q}^{(n)}(\tau) - \langle \mathcal{Q}^{(n)}(\tau)\rangle)^2}
{2(\sigma^{(n)}_{\tau})^2}]$, with $\langle \mathcal{Q}^{(n)}(\tau)\rangle \sim \tau$ and
$\sigma^{(n)}_{\tau}\sim \sqrt \tau$. This form is found with good accuracy, as shown in the
inset of Fig.~\ref{pdfsaboveTc}, where data collapse of the curves with different $\tau $ is
observed by plotting $(\sigma^{(1)}_{\tau})^{1/2}\mathcal{P}(\mathcal{Q}^{(1)}({\tau}))$
against $(\mathcal{Q}^{(1)}(\tau) - \langle
\mathcal{Q}^{(1)}(\tau)\rangle)/\sigma^{(1)}_{\tau}$.

In order to study the  FR (\ref{FR}), we plot the logarithm of the ratio
$\mathcal{P}(\mathcal{Q}^{(n)}(\tau))/ \mathcal{P}(-\mathcal{Q}^{(n)}(\tau))$ as a function of
$\Delta \beta ^{(n)}\mathcal{Q}^{(n)}(\tau)$, see  Fig.~\ref{rette} (inset). For
every value of $\tau $ the data are well consistent with a linear relationship (however, for
large values of the heat the statistics becomes poor), in agreement with the gaussian form of
the PDs. To verify the  FR (\ref{FR}), the slopes of the plot
\begin{equation}
D^{(n)}(\tau)= \frac{\ln{\frac{\mathcal{P}(\mathcal{Q}^{(n)}(\tau))}
{\mathcal{P}(-\mathcal{Q}^{(n)}(\tau))}}}{ \mathcal{Q}^{(n)}(\tau) \Delta \beta ^{(n)}}
\label{sloppe}
\end{equation}
must tend to 1 when $\tau \to \infty$. We show in Fig.\ref{rette} the behavior of the {\it
distance} $\epsilon^{(n)}=1-D^{(n)}(\tau)$ from the asymptotic behavior, for the case with
static coupling to the baths. This quantity indeed goes to zero for large $\tau $ (the same is
found for dynamic couplings). $\epsilon^{(n)}$  depends in general  on the temperatures, the
geometry of the system and on $\tau $. Its behavior can be estimated
on the basis of the following argument.

In systems as the ones considered in this Letter, where generalized detail balance
\cite{BD} holds, the ratio between the probability of a trajectory in configuration space,
given a certain intial condition,
and its time reversed reads
\begin{equation}
\frac{\mathcal{P} (traj)}{\mathcal{P}(-traj)} = e^{\mathcal{Q}^{(n)}(\tau) \Delta \beta ^{(n)}
-\frac{\Delta E}{T_{n'}}},
\label{microrev}
\end{equation}
where $\Delta E = \mathcal{Q}^{(1)}(\tau) + \mathcal{Q}^{(2)}(\tau)$ is the difference
between the energies of the final
and initial state, and $n'\neq n$.
For systems with bounded energy, Eq.~(\ref{microrev}) is the starting point \cite{BD} for obtaining
the FR (\ref{FR}) in the large-$\tau$ limit. Since ${\mathcal Q}^{(n)}(\tau) $ increases with
$\tau $ while $\Delta E$ is limited, in fact, the latter can be asymptotically neglected
and, after averaging over the trajectories, the FR (\ref{FR}) is recovered.
Keeping $\tau $ finite, instead, from Eqs.~(\ref{sloppe},\ref{microrev}) one has that
for the considered trajectory the {\it distance} of the slope from the asymptotic
value is
\begin{equation}
\epsilon ^{(n)}(traj)\simeq \frac{\Delta E} {T_{n'} \mathcal{Q}^{(n)}(\tau )
\Delta \beta ^{(n)}}.
\label{distance}
\end{equation}
We now assume that the behavior of $\epsilon ^{(n)}$ can be inferred by replacing
$\Delta E$ and $\mathcal{Q}^{(n)}(\tau )$ with their average values whose behavior
can be estimated by scaling arguments.
Starting with the average of $\mathcal{Q}^{(n)}(\tau )$, we argue that this quantity
is proportional to $N_{flux} \tau$, $N_{flux} $ being the number of
couples of nearest neighbor spins interacting with baths at different temperatures.
This is because between these spins a neat heat flux occurs.
In the
model with static coupling to the baths one has $N_{flux}\propto L$. With dynamic coupling,
instead, since every spin in the system can feel one or the other temperature, one has
$N_{flux}\propto N$. Concerning the average of
$\Delta E$, this is an extensive quantity proportional to the
number $N$ of spins.  We then arrive at
\begin{equation}
   \epsilon ^{(n)}\simeq (T_{n'} \Delta \beta ^{(n)})^{-1}\, \left \{ \begin{array}{ll}
        L\tau ^{-1}  \qquad $static coupling to baths$  \\
        \tau ^{-1}  \qquad $dynamic coupling to baths.$
        \end{array}
        \right .
        \label{scaling}
\end{equation}
This result is expected to apply for sufficiently large $\tau $ when our scaling approach
holds.

We remind that finite-$\tau $ corrections have been shown to be of order $1/\tau$
for the classes of dynamical systems
considered in \cite{GC95,MR03}.  The same is found in
\cite{Zon} for models based on a Langevin equation \cite{nota},
in cases corresponding to the experimental
setup of a resistor in parallel with a capacitor \cite{cili}.
Faster decays ($\sim 1/\tau^2 $) have been  predicted for other topologies of circuits
\cite{Zon}. On the other hand,
FRs in transient regimes, which are not considered in this
Communication, are exact at any $\tau$ ($\epsilon = 0, \forall \tau$)
\cite{Jarzynski00}.

In our model with static couplings the data of Fig.~\ref{rette} confirm the prediction
of the argument above: curves with different $L$ collapse when plotted against $x=\tau /L$
and $\epsilon ^{(n)}\propto x^{-1}$ for sufficiently large $\tau $. Similar behaviors have
been found by varying the geometry (we also considered rectangular lattices with $L > M$),
transition rates,  and dynamics. The scaling prediction~(\ref{scaling}) has been  verified for
the Ising model with Kawasaki dynamics and squared geometry with $L=10,20,40$, and in  the
case of the system dynamically coupled to the heat baths  for sizes between $L=6$ and $L=60$.

\vspace{0.5cm}

$\mathbf {T_1,T_2 < T_c}$

Let us first recall the behavior of the Ising model
 in contact with  a single bath at
temperature $T<T_c$. When $N=\infty $ the system is confined into one of the two pure states
which can be distinguished by the sign of the magnetization $m(T)=(1/N)\sum _{i=1}^N \sigma
_i$. This state is characterized by a microscopic relaxation time $\tau _{eq}(T)$ which is
related to the fast flip of correlated spins into thermal islands with the typical size of the
coherence length. At finite $N$, instead, genuine ergodicity breaking does not occur. The
system still remains trapped into the basin of attraction of the pure states but only for a
finite ergodic time $\tau _{erg}(N,T)$, which diverges when $N\to \infty$ or $T\to 0$. Then,
as compared to the case $T>T_c$, there is the additional timescale $\tau _{erg}(N,T)$, beside
$\tau _{eq}(T)$, which can become macroscopic. This whole phenomenology is reflected by the
behavior of the autocorrelation function $C(t-t')=\langle \sum _i \sigma _i (t') \sigma _i
(t)\rangle $. When, for large $N$, the two timescales  $\tau _{eq}(T),\tau _{erg}(N,T)$ are
well separated, it first decays from $C(0)=1$ to a plateau $C(t-t')=m(T)^2$ on times
$t-t'\simeq \tau _{eq}(T)$, due to the fast decorrelation of spins in thermal islands. The
later decay from the plateau to zero, observed on a much larger timescale $t-t'\simeq \tau
_{erg}(N,T)$, signals the recovery of ergodicity. Notice also that, from the behavior of
$C(t-t')$ both the characteristic timescales can be extracted.

The same picture applies qualitatively to the
case of two subsystems in contact with two thermal baths, where each systems is trapped
in states with broken symmetry for a time $\tau
_{erg}(N,T_n)$ ($\tau _{erg}(N,T_1)>\tau
_{erg}(N,T_2)$ since $T_1<T_2$),  which can be
evaluated from the autocorrelation functions $C^{(n)}(t-t')=\langle \sum _i
\sigma _i^{(n)} (t)\sigma _i^{(n)}(t') \rangle$, where $\sigma _i^{(n)}$ denote spins in
contact with the bath at $T=T_n$. Since the FR is expected for $\tau $ larger than the
typical timescales of the system, it is interesting to study the role of the additional
timescales $\tau _{erg}(N,T_n)$ on the FR. By varying
$T_1,T_2$ and $N$ appropriately one can realize the limit of large $\tau $ in the two cases
with i) $\tau \ll \tau _{erg}(N,T_2)$ or ii) $\tau \gg \tau _{erg}(N,T_1)$. In case i), in the
observation time-window $\tau $, the system is practically confined into broken symmetry
states while in
case ii) {\it ergodicity} is restored. Not surprisingly, in the latter case, we have observed a
behavior very similar to that with $T_1,T_2>T_c$. The PDs for the case i) are shown in the
lower panel of Fig.~\ref{pdfsaboveTc}. The distributions are more narrow and non Gaussian.
We calculated the skewness of the
distributions defined as
$ <(\mathcal{Q}(\tau) - <\mathcal{Q}(\tau)>)^3>/ <(\mathcal{Q}(\tau)
- <\mathcal{Q}(\tau)>)^2>^{3/2}$.
This quantity is zero  for gaussian distributions while for the case of the lower panel of
Fig.~\ref{pdfsaboveTc} we found   values different from zero which have been
reported in the caption of the figure. This data show that
the PDs slowly approach a Gaussian form increasing $\tau$.
Moreover, differently than in the high
temperature case, an asymmetry between the distributions of  $\mathcal{Q}^{(1)}(\tau)$ and
$\mathcal{Q}^{(2)}(\tau)$ can be observed.

Regarding the slopes $D^{(n)}(\tau)$, they  converge to 1 also in this case even if the times
required to reach the asymptotic behavior are much longer than in the high temperature case
(but always smaller than $\tau _{erg}(N,T_2)$). This suggests that the FR~(\ref{FR}) holds
even in states where ergodicity is broken and that the presence of the macroscopic timescales
$\tau _{erg}(N,T_n)$ does not affect the validity of the FR in this system. Regarding the
scaling of $\epsilon ^{(n)}$, the data are much more noisy than in the case $T_1,T_2>T_c$,
particularly for $\epsilon ^{(1)}$. Despite this, the data presented in the lower
panel of Fig.\ref{rette} for $\epsilon ^{(2)}$ are consistent with the
scaling~(\ref{scaling}) suggesting that it is correct also in this situation.

 In this Letter we have considered different realizations of Ising models coupled to two heat
baths at different temperature. We studied the fluctuation behavior of the heat exchanged with
the thermostats and found that the FR (1) is asymptotically verified in all the cases
considered. We also analyzed  the effects of finite time corrections and their scaling
behavior. The picture is qualitatively similar in the high and low temperature phases,
although very different timescales are required to observe the asymptotic FR.

\acknowledgments

The authors are grateful to A. Pelizzola and L. Rondoni    for useful discussions.


\begin{thebibliography}{99}
\bibitem{nota} In the system studied in \cite{Zon}, $1/\tau$ corrections have been
predicted for any range of work fluctations and only  for some range  of heat fluctuations.
Moreover, differently than in our case,
the slope 1 in work fluctuations is reached from above, i.e. $\epsilon $ is negative.
\end{thebibliography}
\end{document}